\documentstyle[aps10,aps]{revtex}
\begin{document}
\draft
\title{{\hspace*{4.5in} {\rm Liverpool preprint: LTH 374}}\\
{\hspace*{4in} {\it to be published in Physical Review Letters}}\\
On the $\alpha-$decay of deformed actinide nuclei}

\author{T. L. Stewart, M. W. Kermode, D. J. Beachey}
\address{
Theoretical Physics Division, Department of Mathematical Sciences,
University of Liverpool, \\
PO Box 147, Liverpool  L69 3BX, UK}
\author{N. Rowley}
\address{
Centre de Recherches Nucl{\'{e}}aires, 23 Rue du Loess, F 67037 Strasbourg,
CEDEX 2, France}
\author{I. S. Grant}
\address{
Department of Physics and Astronomy,
University of Manchester, Manchester, M13 9PL, UK}
\author{A. T. Kruppa }
\address{
Institute of Nuclear Research of the Hungarian Academy of Sciences,
H-4001 Debrecen, Pf. 51, Hungary}

\date{May 22, 1996}

\maketitle

\begin{abstract}
 $\alpha-$decay through a deformed potential barrier produces
 significant mixing of angular momenta when mapped from the
 nuclear interior to the outside.  
 Using experimental branching ratios and either semi-classical or
 coupled-channels transmission matrices, we have found that
 there is a set of internal amplitudes which are essentially
 constant for all even--even actinide nuclei.
 These same amplitudes also give good results
 for the known anisotropic $\alpha-$particle emission of the favored 
 decays of odd nuclei in the same mass region.
\end{abstract}


The phenomenon of ``tunneling in the presence of an environment" is
of considerable interest in many branches of physics and chemistry
\cite{tokyo}, for example in the tunneling of Cooper pairs through a
Josephson junction\cite{jose}.  In the domain of nuclear physics, the
study of the sub-barrier fusion of heavy nuclei has made significant
contributions to this problem over recent years, in that the
experimental distribution of fusion barriers has been shown to be
intimately related to the environment consisting of the target and
projectile excited states\cite{fusion}.  In this problem the incident
wave is known and all the transmitted flux ends up in the single fusion
channel.  The phenomenon of $\alpha-$decay is potentially more
difficult, since the incident wave ($\alpha-$particle wave function in
the nuclear interior) is unknown but potentially much more rewarding
since (a) for even--even nuclei the transmitted flux may end up in
different daughter states for which the individual fluxes, i.e. 
branching ratios, can
be measured and (b) since for odd nuclei, the presence of different
angular momenta in each daughter state may lead to a 
measurable anisotropy in the $\alpha-$emission.

The $\alpha-$decay of deformed nuclei may be divided into two
distinct parts: the formation of an $\alpha-$particle in the nuclear
interior, followed by its penetration through the $\alpha-$daughter
deformed Coulomb barrier.  There are various approaches to the formation
problem.
One of these assumes a preformed $\alpha-$particle (or at least a
spatially correlated 4--nucleon cluster with the appropriate quantum
numbers), which moves in the deformed field of the daughter nucleus
\cite{preston,clusters}.  Fr\"oman \cite{from57} assumes a constant
$\alpha-$particle probability on the deformed nuclear surface.  Mang
{\em et al.\/} \cite{mang} and, more recently, Delion {\em et al.\/}
\cite{BCS} have considered the deformed single-particle (Nilsson) states
in the vicinity of the Fermi surface and taken the overlap of the
correlated neutron and proton BCS wave functions with an
$\alpha-$particle at the nuclear surface.

Whatever the formation mechanism, the decay proceeds by penetration
through the Coulomb barrier.  If this barrier is assumed to be
spherical, there will be no mixing of orbital angular momentum states
during the tunneling and, for an even--even nucleus, the observed
branching ratios to different rotational states of the daughter ($I=0^+,
2^+, 4^+$, ....) will be determined by the $L-$admixture in the nuclear
interior, modified by the transmission factors for the different
centrifugal barriers evaluated at slightly different $\alpha-$particle
energies due to the excitation energies $\epsilon^*_I$ of the daughter.
If one takes account of the deformation of the barrier, there will be
additional mixing during the tunneling \cite{from57,bohr75}.
Such effects have been considered in the fusion of heavy nuclei 
\cite{lipp83} and more recently confirmed experimentally \cite{wei91,lei93}.
In this Letter, we take known branching ratios and calculate the mixing
in the barrier to obtain the internal amplitudes
near the nuclear surface. No model of internal dynamics 
($\alpha-$preformation or distribution) is required and indeed one does
not know what preformation factors have to be fitted unless the barrier
penetration problem is first solved. 
We then systematically survey the internal amplitudes for the even--even 
actinide $\alpha-$emitters for various choices of relative phases.
A surprising result that emerges from our analysis is that the relative
internal amplitudes may possibly be constant over a wide range of actinide
nuclei and this in itself gives a very strong indication of the type of
nuclear model for the preformation factor which might be the most appropriate.

The fine structure of the $\alpha-$particle energy spectrum
determines the branching ratios to states of known spins $I$ in the
daughter.  Since, for an even--even nucleus, the orbital angular
momentum of the $\alpha-$particle must be $L=I$, we shall set up our
problem in terms of these outgoing $L-$waves.  In the intrinsic frame of
the axially--symmetric deformed daughter, one may write an asymptotic
outgoing wave in the form
\begin{equation}
\psi^{outside}  =
\sum_L C_L \,{\cal{O}}^{outside}_L(k_L,{\bf r})=
\sum_L C_L \,{\bigl(}G_L(k_L,r)+iF_L(k_L,r)\bigr{)}Y_{L0}
(\hat{\bf r})\chi_{L0}, \label{e2}
\end{equation}
where $F_L$ and $G_L$ are the regular and irregular Coulomb wave
functions.  The ${\cal{O}}^{outside}_L$ represent outgoing
$\alpha-$particles with orbital angular momentum $L$ coupled to total
angular momentum zero with a state $\chi_{L0}$ of the daughter of spin
$I=L$.  The wave numbers $k_L$ differ due to the different
$\epsilon^*_I$.  With such an outgoing wave, one can perform a
coupled--channels calculation to obtain the wave function on the other
side of the barrier.  This must have both incoming and outgoing
components
\begin{equation}
\psi^{inside}=\sum_L \left(A_L \,{\cal{O}}^{inside}_L(k_L,{\bf r})+B_L\,
{\cal{I}}^{inside}_L(k_L,{\bf r})\right).
\end{equation}
The corresponding currents are shown schematically in Fig. 1.

\begin{figure}[ht]
\vspace{80mm}
\includegraphics{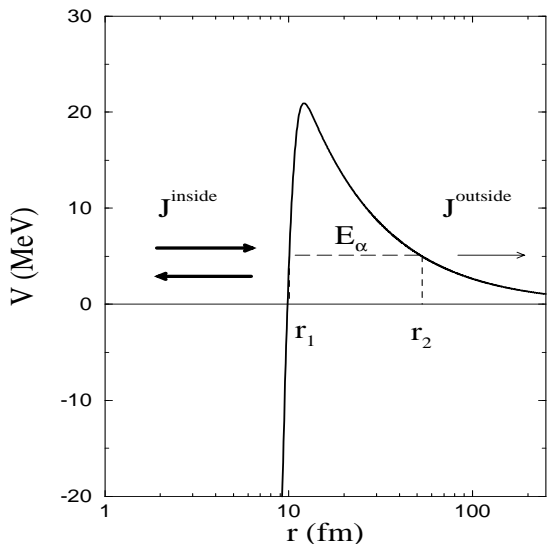}
\caption{Schematic diagram of the potential V and the incoming and
outgoing currents. The classical turning points $r_1$ and $r_2$, and the
energy of the $\alpha-$particle $E_{\alpha}$ are used in the
semi--classical calculation of the transmission matrix, see Eq. (\ref{e5}).}
\end{figure}

The above problem may be solved numerically to obtain the outside
coefficients $C_L$ in terms of those inside, $A_{L}$. One may thus
define a transmission matrix $M$
\begin{equation}
C_L=\sum_{L'}M_{L\,L'}\,A_{L'}. \label{e1}
\end{equation}
Since we wish to undertake a systematic study of $\alpha-$emitters
in the actinide region, we shall first obtain the transmission matrix
using the semi--classical method outlined below.  We thus express the
outer amplitudes $c_L$ (lower case indicates WKB solutions) in terms of
the amplitudes $a_L$ in the nuclear interior by
\begin{equation}
c_L = \sum_{L'}{\cal K}_{LL'} a_{L'}
=\sum_{L'} \langle Y_{L0} \vert \exp(-I_L(\cos \theta)) \vert
Y_{L'0}\rangle a_{L'}
\label{e4}
\end{equation}
where $I$ takes the usual WKB form
\begin{equation}
I_L(\cos \theta) = \int_{r_1(\cos \theta)}^{r_2}
\biggl[ {{2\mu }\over{\hbar^2}}(V_L(r,\cos\theta)-E_{\alpha})
\biggr]^{1\over 2} {\rm d}r\,,
\label{e5}
\end{equation}
(see Fig. 1). The potential $V_L(r,\cos\theta)$ comprises the Coulomb
field due to the deformed charge distribution of the daughter, a
deformed Woods-Saxon potential and a centrifugal term. We consider both
$\beta_2$ and $\beta_4$ terms in the deformation.
The angular integrals were evaluated using the technique of Kermode and
Rowley \cite{kerm93}.
The matrix ${\cal K}_{LL'}$ is the analogue of $M_{LL'}$ in the
coupled-channels formalism.

The above approach is essentially an exact calculation of the
semi--classical transmission coefficients introduced by Fr\"oman
\cite{from57} and used, for example, in Ref. \cite{BCS}.  Indeed the
earlier expressions of Fr\"oman may be obtained from Eq. (\ref{e4}) by
ignoring the hexadecapole deformation, taking the nuclear potential to
be a $\theta-$dependent sharp cut--off and by making a first--order
expansion in $\beta_2$.  Approximations similar to these have also been
employed in a coupled--channels formalism \cite{ras56,pen58}.

The magnitudes of the coefficients $c_L$ are proportional to the
square roots of the branching ratios for the angular momenta $L$.  They
may be taken to be real and can, in principle, be either positive or
negative.  Their reality follows from the requirement that the imaginary
part of the wave function at the nuclear surface should be small
\cite{preston}.
In this Letter, we restrict our calculations to $\{L\}=\{0,2,4\}$
since few nuclei have measured branches to the 6$^+$ state.  (We have,
however, considered the inclusion of 6$^+$ and $8^+$states and have
found that our conclusions are essentially unaffected.)  We consider 4
possibilities for the relative signs of the $\{c_L\}$, i.e. $\{+++\}$,
$\{++-\}$, $\{+-+\}$ and $\{+--\}$.  For all the decays we consider, the
Sommerfeld parameters are large and the Coulomb phases then ensure that
for the case $\{+++\}$ the spherical harmonics in Eq. (\ref{e2}) are in
phase along the symmetry axis.  Thus if one considered the amplitudes in
Eq. (\ref{e2}) to add coherently, the outgoing flux would be axial.  In
the even--even system, of course, the currents are not added coherently
and the outgoing flux is always isotropic for each $L$, after
integrating over all orientations of the daughter.  However, we shall
see below that the above consideration is important for odd--even
systems which may be polarized to yield anisotropic $\alpha-$decay.  For
each of the above phase combinations, we have determined the amplitudes
$a_L$ from Eq. (\ref{e4}) using deformation parameters from M\"oller
{\em et al.\/} \cite{moll93}.  Figure 2 shows the 4 sets of $a_L$
(normalised to unity) for actinide nuclei with daughter mass (atomic
number) $220\le A_d\le 248$ ($88\le Z_d\le 96$).

\begin{figure}[ht]
\vspace{85mm}
\includegraphics{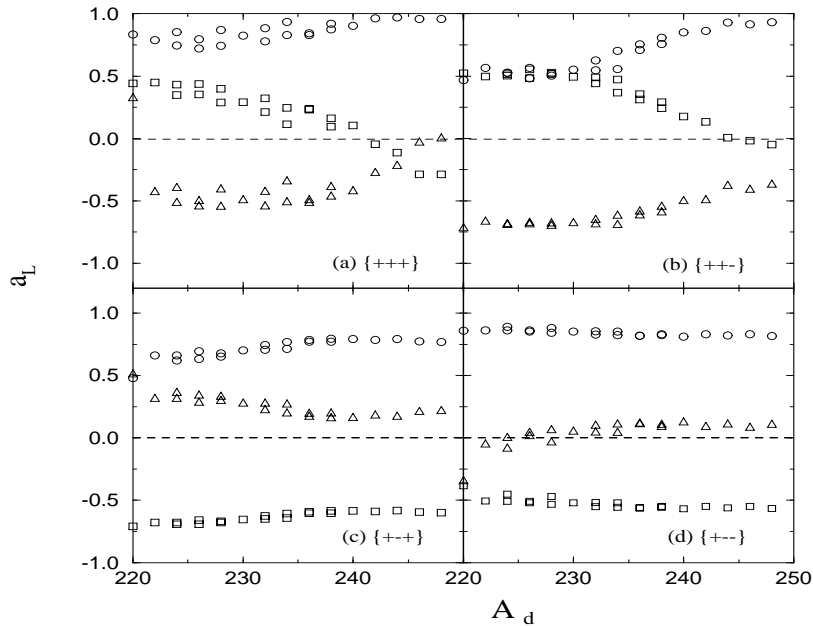}
\caption{The amplitudes $a_L$ for even--even actinide nuclei with
the four choices of external phase indicated.
The circles, squares and triangles represent $L=0,~L=2~{\rm and}~L=4$,
respectively.}
\end{figure}

To test the accuracy of the WKB approximation, we have performed exact
coupled--channels calculations for three of the above nuclei.  The
results were in good agreement with those shown in this figure.  For
example, for $^{238}$U with the phase choice $\{c_L\} = \{+--\}$, the
coupled--channels give $\{a_L\} = \{0.84, -0.54, 0.08\}$ compared with
the semi--classical values $\{a_L\} = \{0.83, -0.55, 0.10\}$.  The
importance of the $L-$mixing under the barrier is demonstrated by the
corresponding results $\{a_L\} = \{ 0.70, -0.68,-0.19\}$ for the {\it
spherical} case.

Since the deformation of the daughter varies smoothly with $A_d$,
one might expect that just one set of the solutions shown in Fig. 2
corresponds to the physical amplitudes.  One particularly interesting
solution is that obtained with the combination $\{+--\}$ (Fig. 2d) since
the coefficients $a_L$ are practically nucleus--independent even though
$\beta_2$ varies from around 0.10 to 0.24 and the $\alpha-$particle
energies vary from around 4 to 7 MeV over this mass region.  It is also
the only one to yield amplitudes with $\vert a_0 \vert > \vert a_2\vert
> \vert a_4\vert $ for all nuclei (see the discussion below).

We note, however, that for the constant solution $a_2$ is always negative.
At first sight, this implies that if the same
solution prevailed for odd--even nuclei, then $\alpha-$emission 
takes place mainly in the equatorial plane -- a result which would
not be consistent with measured anisotropies.  
Consider, however, the
favored $\alpha-$decay of an aligned nucleus with 
$ K=M= J_p = J_d$, where $K$ is the spin projection of the
nucleus along the symmetry axis, $M$ the projection of the parent
spin $J_p$ along the space-fixed $z-$axis and $J_d$ the daughter spin.
For this case the angular distribution of the emitted $\alpha-$particles 
may be written (see, for example, \cite{from57,kerm93})
\begin{equation}
W(\theta) \propto
\sum_m \Big| \sum_L \langle J_d~L~M-m~m \vert J_p~M\rangle
\langle J_p~L~K~0 \vert J_d~K\rangle~c_L~Y_{Lm}(\theta,0)\Big| ^2 ,
\label{e7}
\end{equation}
where the Clebsch-Gordon coefficients arise out of the transformation 
from the body-fixed to the lab frame.
We define the anisotropy by  $W(0)/W({1\over2}\pi)$.
Equation (\ref{e7}) describes the favored decay $K_p=K_d=K$ angular
distribution only. A contribution from the decay $K_p=K\rightarrow K_d=-K$
is suppressed (see, for example, page 50 of \cite{from57} or 
pages 272-3 of \cite{bohr75}).

A simple model for the odd--even nucleus would be to assume that it
consists of an even--even core plus a spectator nucleon.  Indeed Fr\"oman
\cite{from57} employed the above formula using $c_L$ extracted from the
neighbouring even--even branching ratios, which implicitly uses the same
transformation (\ref{e4}) as for the even--even case.  Delion {\em et
al.\/} \cite{BCS} have also used Eq. (\ref{e4}) but derive their
amplitudes $a_L$ from a BCS calculation.  The expression (\ref{e4}) may
not, however, be appropriate for odd--even systems, where the
$\alpha-$particle energy is determined by the daughter spin $J_d$
rather than by the orbital angular momentum $L$.  We have applied Eq.
(\ref{e4}) to the four odd--even actinide nuclei for which anisotropies
have been measured (see Table ~\ref{t0}), using the amplitudes $a_L$
from Fig. 2.  We find that Eq. (\ref{e4}) cannot predict anisotropies
both greater than and less than one for these nuclei, for any of the
four sets of $a_L$.  This is because the sign of $c_2/c_0$ is either
positive (anisotropy $>1$, i.e.  Figs. 2a,2b) or negative (anisotropy
$<1$, i.e.  Figs. 2c,2d).

If the energies of the excited states of the daughter were high,
then barrier penetration would filter out the components of the
wave function corresponding to configurations other than its ground
state.  
In that case we obtain
\begin{equation}
c_L  \approx \sum_{L'}{\cal K}_{LL'}
\langle J_p~L'~K~0 \vert J_d~K\rangle^2 a_{L'}. \label{e10}
\end{equation}
We note that Bohr and Mottelson \cite{bohr75} suggest a 
similar $J_d$-decoupled equation for the leading order transition rates.
Equation (\ref{e10}) has also been applied to the four nuclei in
Table~\ref{t0}, using the internal amplitudes of Fig. 2d.

\begin{table}
\caption{Anisotropies ($W(0)/W({1\over 2}\pi)$), for the 
favored decay of four odd--even nuclei.
Deformation parameters correspond to the daughter nucleus.}
\begin{tabular}{lcllccc}
Parent & $J_p$  & $\beta_2$ & $\beta_4$ &  Theory $\{+--\}$
& Experiment & ref.   \\ \hline \\
$^{221}$Fr & ${5\over 2}$  & 0.039 $^*$ & 0.028 &  0.77 & 0.37(2)
& \cite{sch94}      \\
$^{227}$Pa & ${5\over 2}$  & 0.147 $^*$  & 0.110 &  2.66 & 3.55(28)
& \cite{sch94}      \\
$^{241}$Am & ${5\over 2}$ & 0.215  & 0.102 & 4.26 & $>$2.7
& \cite{amer}       \\
$^{253}$Es$^{**}$ & ${7\over 2}$  & 0.235  & 0.040 & 3.70 & $>$3.8
& \cite{es253}
\end{tabular}
$^*$ {\small M\"oller {\em et al.\/} give a non-zero value for
$\beta_3$}

$^{**}$ {\small includes L=6}

\label{t0}
\end{table}
\vskip 0.2truecm

The values of the predicted anisotropies agree well with the
experimental data.  In particular, we emphasize the prediction of
anisotropies both less than and greater than 1, i.e. it is possible to
have more $\alpha-$particles emitted along the symmetry axis than
equatorially, despite the fact that $a_2<0$.  The reason is that the
external amplitude $c_2$ may become positive, since the Clebsch--Gordan
coefficients of Eq. (\ref{e10}) attenuate the effect of the internal
amplitude $a_2$.  We note that this effect is not possible if we use the
amplitudes from Figs. 2a or 2b.  The solutions of Poggenburg {\em et
al.\/} and of Delion {\em et al.\/} give anisotropies always greater
than one.  However, the excited states of the daughter lie at relatively
low excitation energies and, in the present model, their coupling to the
ground state is not sufficiently attenuated by barrier penetration for
the reversal of sign between $a_2$ and $c_2$ to take place.  The success
of Eq. (\ref{e10}), could, however, suggest some other dynamical element
through which the $\alpha-$particle orbital angular momenta are mixed in
exactly the same way as for even--even nuclei, but the coupling to
different daughter states is absent.

\vfill \eject

We have shown that it is possible to describe all known branching
ratios of even--even actinide nuclei with an $\alpha-$particle
wave function near the nuclear surface which is practically 
nucleus-independent.  
This model has
a certain aesthetic appeal in itself and, with the 
assumption that the same $L$-mixing matrix ${\cal K}_{LL'}$ 
is present in the favored decay of an odd system, is capable of reproducing
the known anisotropies in this mass region.  This of course begs the
question as to what physical model could generate such constant
amplitudes.  The best candidate would appear to be the notion that the
$\alpha-$particle amplitudes should be projected from the
pair-correlated neutron and proton Nilsson-model states.  In this mass
region, the level density is high and the pair forces lead to a rather
diffuse Fermi surface.  One might then expect that the correlated
ground-state wave function should vary rather slowly with the Fermi
energy, or in other words with the nucleon numbers of the system.  Since
pairing mainly takes place through the two-body angular momentum $J=0$,
this model would also be expected to give amplitudes with the property
$\vert a_0 \vert > \vert a_2\vert > \vert a_4\vert $, as found in Fig.
2d.

\vskip 0.2truecm
We thank Paul Schuurmans for helpful comments and for allowing us to
use data prior to publication.
The award of EPSRC grant GR/J21507 is gratefully acknowledged.
NR and ATK are grateful for support through the Royal Society
Collaborative Grant Scheme.
ATK is grateful for support from the Hungarian OTKA Grant No. T17298.


\begin{references}

\bibitem{tokyo}
See {\it e.g.} Proc. 4th Int. Symp. on the Foundations of Quantum
Mechanics, Tokyo, 1992, eds.  M. Tsukada, S. Kobayashi, S. Kurihara and
S. Nomura, JJAP Series 9 (1993) (dedicated to the problem of quantal
tunneling).

\bibitem{jose}
A. J. Leggett, {\it ibid}, p. 10.

\bibitem{fusion}
A. M. Stefanini {\em et al.\/}, Phys.  Rev. Lett. {\bf 74} (1995) 864
and references therein.

\bibitem{preston}
M. A. Preston and R. K. Bhaduri,
{\it Structure of the nucleus}
(Addison-Wesley, Massachusetts, 1975).

\bibitem{clusters}
T. Berggren, Hyperfine Interactions {\bf 75} {(1992)} {401};
N. Rowley, G. D. Jones and M. W. Kermode, Journal of Physics G {\bf 18}
{(1992)} {165};
B. Buck, A. C. Merchant and S. M. Perez, Phys. Rev. C {\bf 45} {(1992)}
{2247}.

\bibitem{from57}
P. O. Fr\"oman, Mat. Fys. Skr. Dan. Vid. Selsk. {\bf 1} (1957) 1.

\bibitem{mang}
H. J. Mang, Ann. Rev. Nucl. Sci. {\bf 14} (1964) 1;
J. K. Poggenburg, H. J. Mang and J. O. Rasmussen, Phys. Rev. {\bf 181}
(1969) {1697}.

\bibitem{BCS}
D. S. Delion, A. Insolia and R. J. Liotta, Phys. Rev. C {\bf 46}
{(1992)} 884;
D. S. Delion, A. Insolia and R. J. Liotta, Phys. Rev. C {\bf 46}
{(1992)} 1346;
D. S. Delion, A. Insolia and R. J. Liotta, Phys. Rev. C {\bf 49}
{(1994)} 3024.

\bibitem{bohr75}
A. Bohr and B. R. Mottelson,
{\it Nuclear Structure Vol. I and II}
(Benjamin, New York, 1975).

\bibitem{lipp83}
R. Lipperheide, H. Rossner and H. Massmann, Nucl. Phys. {\bf A394} (1983)
312.

\bibitem{wei91}
J. X. Wei, J. R. Leigh, D. J. Hinde, J. O. Newton, R. C. Lemmon, 
S. Elfstr\"om, J. X. Chen and N. Rowley, 
Phys. Rev. Lett. {\bf 67} (1991) 3368.

\bibitem{lei93}
J. R. Leigh, N. Rowley, R. C. Lemmon, D. J. Hinde, J. O. Newton, 
J. X. Wei, J. Mein, C. Morton, S. Kuyucak and A. T. Kruppa, Phys. Rev. C
{\bf 47} (1993) R437.

\bibitem {kerm93}
M. W. Kermode and N. Rowley, Phys. Rev. C {\bf 48} (1993) {2326}.

\bibitem{ras56}
J. O. Rasmussen and B. Segall, Phys. Rev. {\bf 103} (1953) 1102.

\bibitem{pen58}
E. M. Pennington and M. A. Preston, Can. J. Phys. {\bf 36} (1958) 944.

\bibitem {moll93}
P. M\"oller, J. R. Nix, W. D. Myers, W. J. Swiatecki,
Atomic Data and Nuclear Data Tables {\bf 59} (1995) 185.

\bibitem{sch94}
P. Schuurmans, Leuven ({\it private communication}).

\bibitem{amer}
A. J. Soinski and D. A. Shirley, Phys. Rev. C {\bf 10} (1974) {1488}.

\bibitem{es253}
A. J. Soinski and D. A. Shirley, Phys. Rev. C {\bf 2} (1970) {2379}.

\end{references}
\end{document}